\documentclass[11pt]{article}
\usepackage{amsmath,amsthm,amssymb,cite,enumerate}
\usepackage[a4paper,left=0.5in,right=1in]{geometry}
\usepackage[usenames]{color}
\usepackage{graphicx}
\usepackage{epsfig}
\usepackage{dcolumn}% Align table columns on decimal point
\usepackage{bm}% bold math

\usepackage{authblk} %for authors with multiple affiliations

\begin{document}

\def \d {{\rm d}}

\def \bm #1 {\mbox{\boldmath{$m_{(#1)}$}}}

\def \bF {\mbox{\boldmath{$F$}}}
\def \bV {\mbox{\boldmath{$V$}}}
\def \bff {\mbox{\boldmath{$f$}}}
\def \bT {\mbox{\boldmath{$T$}}}
\def \bk {\mbox{\boldmath{$k$}}}
\def \bl {\mbox{\boldmath{$\ell$}}}
\def \bn {\mbox{\boldmath{$n$}}}
\def \bbm {\mbox{\boldmath{$m$}}}
\def \tbbm {\mbox{\boldmath{$\bar m$}}}

\def \T {\bigtriangleup}
\newcommand{\msub}[2]{m^{(#1)}_{#2}}
\newcommand{\msup}[2]{m_{(#1)}^{#2}}

\newcommand{\be}{\begin{equation}}
\newcommand{\ee}{\end{equation}}

\newcommand{\beqn}{\begin{eqnarray}}
\newcommand{\eeqn}{\end{eqnarray}}
\newcommand{\AdS}{anti--de~Sitter }
\newcommand{\AAdS}{\mbox{(anti--)}de~Sitter }
\newcommand{\AAN}{\mbox{(anti--)}Nariai }
\newcommand{\AS}{Aichelburg-Sexl }
\newcommand{\pa}{\partial}
\newcommand{\pp}{{\it pp\,}-}
\newcommand{\ba}{\begin{array}}
\newcommand{\ea}{\end{array}}

\newcommand{\M}[3] {{\stackrel{#1}{M}}_{{#2}{#3}}}
\newcommand{\m}[3] {{\stackrel{\hspace{.3cm}#1}{m}}_{\!{#2}{#3}}\,}

\newcommand{\tr}{\textcolor{red}}
\newcommand{\tb}{\textcolor{blue}}
\newcommand{\tg}{\textcolor{green}}

\def\a{\alpha}
\def\g{\gamma}
\def\de{\delta}

\def\b{{\kappa_0}}

\def\E{{\cal E}}
\def\B{{\cal B}}
\def\R{{\cal R}}
\def\F{{\cal F}}
\def\L{{\cal L}}

\def\e{e}
\def\bb{b}

\title{Electromagnetic fields with vanishing quantum corrections}

\author[1,2]{Marcello Ortaggio\thanks{ortaggio(at)math(dot)cas(dot)cz}}
\author[1]{Vojt\v ech Pravda\thanks{pravda@math.cas.cz}}

\affil[1]{Institute of Mathematics of the Czech Academy of Sciences, \newline \v Zitn\' a 25, 115 67 Prague 1, Czech Republic}
\affil[2]{Instituto de Ciencias F\'{\i}sicas y Matem\'aticas, Universidad Austral de Chile, \newline Edificio Emilio Pugin, cuarto piso, Campus Isla Teja, Valdivia, Chile}

\maketitle

%\date{\today}

\begin{abstract}
We show that a large class of null electromagnetic fields are immune to any modifications of Maxwell's equations in the form of arbitrary powers and derivatives of the field strength. These are thus exact solutions to virtually any generalized classical electrodynamics containing both non-linear terms and higher derivatives, including, e.g., non-linear electrodynamics as well as QED- and string-motivated effective theories. This result holds not only in a flat or (anti-)de~Sitter background, but also in a larger subset of Kundt spacetimes, which allow for the presence of aligned gravitational waves and pure radiation.
\end{abstract}

%\pacs{03.50.De, 03.50.Kk, 11.10.Lm, 04.40.Nr}


\section{Introduction}

\label{sec_intro}

Early attempts to modify Maxwell's theory in order to cure the divergent electron's self-energy date back to Mie's work of 1912 \cite{Mie12}. A physically more satisfactory non-linear theory was subsequently proposed by Born and Infeld \cite{Born33,BorInf34}. From a quantum perspective, non-linear deviations from Maxwell's theory also appear in the effective Lagrangian derived by Heisenberg and Euler from quantum electrodynamics (QED), relevant to light-by-light scattering and electron-positron pair production \cite{HeiEul36,Weisskopf36,Schwinger51}. Both the Born-Infeld and Heisenberg-Euler theories are special cases of non-linear electrodynamics (NLE), a more general theory defined by a Lagrangian which depends (in principle arbitrarily) on the two algebraic invariants $I_1\equiv F_{ab}F^{ab}$ and $I_2\equiv F_{ab}*F^{ab}$ \cite{Plebanski70}. The resulting field equations take the form 
\be
 \d \bF =0 , \qquad *\d*\tilde \bF =0 , 
 \label{univ}
\ee
where $\tilde\bF=\alpha\bF+\beta{}^*\bF$ and $\alpha$, $\beta$ are specific scalar functions of $I_1$ and $I_2$. Starting from the 80s, interest in NLE has been revived by the observation that the Born-Infeld Lagrangian emerges as an effective low-energy Lagrangian in string theory (cf., e.g., \cite{Tseytlin00} for a review and references). 

When it comes to finding exact solutions, non-linear theories such as NLE are obviously much more difficult to deal with. 
It thus perhaps came as a surprise when Schr\"odinger pointed out that all {\em null} fields (defined by $I_1=0=I_2$) which solve Maxwell's theory also automatically solve any non-linear electrodynamics (in vacuum) \cite{Schroedinger35,Schroedinger43}. In fact, Schr\"odinger's observation can readily be extended to theories more general than NLE, by letting $\tilde\bF$ in \eqref{univ} be any 2-form constructed algebraically and polynomially from $\bF$. In a nutshell, the proof of this fact boils down to observing that, if $\bF$ is null, any such $\tilde\bF$ can only be a linear combination with constant coefficients of $\bF$ and $*\bF$ (quadratic and higher terms vanish thanks to the antisymmetry of $\bF$), and therefore obeys Maxwell's equations if $\bF$ does (although NLE was originally formulated in a flat spacetime, the same argument holds also in curved backgrounds). 
Null fields are thus solutions displaying theory-independent properties. They are also of interest from a physical viewpoint, since they characterize electromagnetic plane waves \cite{Schwinger51,syngespec} as well as the asymptotic behaviour of radiative systems \cite{penrosebook2}, and they approximate the field produced by high-energy sources \cite{Bergmann42,syngespec,RobRoz84}. Additionally, they are also relevant to Penrose's limits in supergravity \cite{Guven87}.

It would thus be desirable to identify classes of exact solutions possessing similar ``universal'' properties in a context more general than NLE, i.e., including higher derivatives theories. These would naturally have broader physical applications, since effective Lagrangians depending also on the derivatives of the field strength are relevant both to classical \cite{Bopp40,Podolsky42} and quantum \cite{Euler36,Deser75,Dunne05} field theories,  as well as to string theory \cite{Tseytlin00}. However, the simple argument we have presented above breaks down if $\tilde\bF$ contains derivatives $\nabla^{(k)}\bF$, so that Schr\"odinger's observation does not readily extend to higher-derivative electrodynamics.

It is the purpose of the present contribution to show that, nevertheless, a large family of universal Maxwell fields exists that solve any theory defined by a Lagrangian constructed out of $\bF$ and its derivatives of arbitrary order. More precisely, we will identify {\em universal} solutions of the system~\eqref{univ}, where the second equation holds simultaneously for {\em all} $2$-forms $\tilde \bF $ constructed polynomially from $\bF $ and its covariant derivatives (Maxwell's theory of course corresponds to the special choice $\tilde \bF =\bF $). Because $\tilde \bF $ contains derivatives, in order to be able to prove this fact we will need to place some restrictions also on the background spacetime. However, the most standard case of flat, de~Sitter (dS) and anti-de~Sitter (AdS) spacetimes will be included in our result. Throughout the paper, we will restrict ourselves to the (modified) sourcefree Maxwell equations in a four-dimensional spacetime. Some extensions and applications will be mentioned in the final Discussion.

\section{The solutions}

\label{sec_solutions}

Technically, our result may be stated as follows: {\em in a Kundt spacetime of Petrov type III and aligned traceless-Ricci type III, any aligned null $\bF$ that solves Maxwell's equations is universal}. More explicitly, this means that the spacetime can be written as
\be
 \d s^2=2P^{-2}\d\zeta\d\bar\zeta-2\d u\left(\d r+W\d\zeta+\bar W\d\bar\zeta+H\d u\right) ,
\label{metric}
\ee
where 
\beqn
  & & P=P(u,\zeta,\bar\zeta) , \\
  & & W=rg^{(1)}_{,\zeta} +g^{(0)}(u,\zeta,\bar\zeta) , \\
	& & H=r^2H^{(2)}(u,\zeta,\bar\zeta)+rH^{(1)}(u,\zeta,\bar\zeta)+H^{(0)}(u,\zeta,\bar\zeta) .
\eeqn
The function $g^{(0)}$ is complex and arbitrary, $g^{(1)}=g^{(1)}(u,\zeta,\bar\zeta)$ and all the remaining functions are real with $H^{(1)}$ and $H^{(0)}$ arbitrary. Further constraints following from the type III curvature conditions are
\beqn
	& & \Delta\ln P=K(u) , \label{K} \\
  & & 2H^{(2)}=-\frac{1}{4}P^2\left(g^{(1)}_{,\zeta}g^{(1)}_{,\bar\zeta}+2g^{(1)}_{,\zeta\bar\zeta}\right) , \label{H2} \\
	& & \Delta\ln P+\frac{1}{4}P^2\left(g^{(1)}_{,\zeta}g^{(1)}_{,\bar\zeta}-2g^{(1)}_{,\zeta\bar\zeta}\right)=0 , \label{DeltaP} \\
	& & \left(P^2g^{(1)}_{,\zeta}\right)_{,\zeta}-\frac{1}{2}\left(Pg^{(1)}_{,\zeta}\right)^2=0 , \label{g1}
\eeqn
where $\Delta\equiv 2P^2\pa_\zeta\pa_{\bar\zeta}$. Then any 2-form of the form
\be
 \bF=\d u\wedge\left[f(u,\zeta)\d\zeta+\bar f(u,\bar\zeta)\d\bar\zeta\right] ,  
 \label{F_N_coords}
\ee
where $f(u,\zeta)$ is an arbitrary complex function, solves identically the equations~\eqref{univ}, with $\tilde \bF $ as described above. The multiple principal null direction (of both the Riemann tensor and of $\bF$) is defined by $\bl=\pa_r$. 

Eq.~\eqref{K} means that, at a given $u$, the 2-space $2P^{-2}\d\zeta\d\bar\zeta$ has constant Gaussian curvature $K(u)$. A redefinition $\zeta\to\zeta'(u,\zeta)$ thus enables one to set (cf., e.g., eq.~(2.55) of \cite{NewTamUnt63} or (7.2) of \cite{Talbot69})
\be
 P=1+\frac{K(u)}{2}\zeta\bar\zeta .
\ee
Using this, one can integrate explicitly \eqref{g1}, which gives (after requiring that $g^{(1)}$ is real and imposing \eqref{DeltaP} with \eqref{K}) \cite{OzsRobRoz85,GriDocPod04}
\be
 g^{(1)}=2\ln\frac{P}{a(u)+b(u)\zeta+\bar b(u)\bar\zeta-a(u)\frac{K(u)}{2}\zeta\bar\zeta} ,
\ee
where $a(u)$ and $b(u)$ are arbitrary functions of $u$ (real and complex, respectively). The invariant sign of $H^{(2)}$ is then determined by the combination $-(b\bar b+a^2\frac{K}{2})$ \cite{OzsRobRoz85,GriDocPod04}. Depending on this sign, a remaining coordinate freedom can be used to cast $g^{(1)}$ into various canonical forms, as detailed in \cite{OzsRobRoz85,GriDocPod04}.\footnote{This part of the analysis of \cite{OzsRobRoz85,GriDocPod04} (performed in the case $K=$const) extends straightforwardly also to the case $K=K(u)$.} 
These spacetimes include, in particular, Minkowski and (A)dS, as well as various gravitational and electromagnetic waves propagating in those backgrounds \cite{Stephanibook,GriPodbook}. For example, particular solutions to \eqref{H2}--\eqref{g1} (with $K=2\Lambda=$const) are given by $P=1+\frac{1}{6}\Lambda\zeta\bar\zeta$, $g^{(1)}_{,\zeta}=2\bar\tau P^{-1}$, $H^{(2)}=-\left(\tau\bar\tau+\frac{1}{6}\Lambda\right)$, with $\tau=\frac{1}{3}\Lambda\zeta\left(1-\frac{1}{6}\Lambda\zeta\bar\zeta\right)^{-1}$ or $\tau=-\left(1-\frac{1}{6}\Lambda\zeta^2\right)\left(\zeta+\bar\zeta\right)^{-1}$ or (for $\Lambda<0$) $\tau=-\sqrt{-\frac{1}{6}\Lambda}\left(1+\sqrt{-\frac{1}{6}\Lambda}\zeta\right)\left(1+\sqrt{-\frac{1}{6}\Lambda}\bar\zeta\right)^{-1}$ -- these include generalized Kundt waves in (A)dS, Siklos waves in AdS and \pp waves \cite{GriPodbook}. Spacetimes for which all the scalar curvature invariants vanish are also comprised in \eqref{metric} \cite{Pravdaetal02}.

\subsection*{Proof}
Let us sketch a proof of the above statement (more technical details will be presented elsewhere along with some generalizations). The first of~\eqref{univ} is obviously satisfied because it takes the same form as in Maxwell's theory. The strategy is thus to prove that, under the above assumptions, all the possible $\tilde\bF$ that one can construct are automatically divergencefree (or zero). 
The Kundt assumption means that the spacetime admits a geodesic null vector field $\bl=\pa_r$ with vanishing expansion, shear and twist, which can be written as
\be
 \ell_{(a;b)}=2\ell_{(a}\xi_{b)} , \qquad \ell^a\xi_b=0 , \qquad \ell_{[a,b}\ell_{c]}=0 .
\ee
 Additionally, the conditions on the curvature mean that the Riemann tensor is of the form,
\be
 R_{abcd}=\frac{R}{6} g_{a[c}{g_{d]b}}+\ldots ,
 \label{riem_spec}
\ee
where the ellipsis denote terms of negative boost-weight, i.e., such that no non-zero scalar invariant can be constructed out of those algebraically (their explicit form is not important here). Thanks to \cite{OrtPra16}, the assumptions ensure that all the scalar invariants constructed out of $\bF $ and $\nabla^{(k)}\bF$ vanish identically. A result of \cite{Hervik11} and the antisymmetry of $\tilde\bF$ imply that $\tilde\bF$ must be a {\em linear} combination with constant coefficients of $\bF$, $*\bF$ and (suitable contractions of) terms $\nabla^{(k)}\bF$ and $\nabla^{(k)}*\bF$ with $k=1,2,\ldots$ (roughly speaking, quadratic and higher terms would contain at least two non-contracted $\bl$s, thus giving zero by antisymmetry). Now, the terms $\bF$ and $*\bF$ are harmless since they are divergencefree by assumption. The remaining terms need $k/2$ contractions with the metric tensor in order to produce an object of rank~2. In flat space covariant derivative commute, so that all possible terms can be rewritten in such a way as to contain (derivatives of) the divergence of $\bF$ or $*\bF,$ and the proof is already complete. In a curved spacetime this is not true, since shifting indices produces terms linear in the curvature tensor, via the generalized Ricci identity (schematically, for a tensor $\bT$ one has $[\nabla,\nabla]\bT=\bT\cdot\mbox{Riem}$ -- cf., e.g., \cite{Waldbook}). However, repeated use of the latter together with the Weitzenb\"ock identity for $p$-forms \cite{Weitzenbock23} leads, thanks to \eqref{riem_spec}, to the same conclusion. Although this idea is straightforward, the intermediate steps are lengthy and require several technicalities (similar to those used in \cite{Pravdaetal02,OrtPra16} and references therein) necessary to prove certain special differential properties of the Riemann tensor that hold in the spacetimes~\eqref{metric} (the simplest example of those properties being $R_{,a}\propto \ell_a$). These enable one, ultimately, to reorder the covariant derivatives in a desired way and thus to show that any $\tilde\bF$ is divergencefree, which completes the proof.

\section{Discussion}

The electromagnetic fields described in this paper represent a large family of universal vacuum solutions, thus being relevant to virtually any classical theory of electrodynamics. They are a subset of {\em null} fields that include, in particular, plane-waves-like solutions propagating in various backgrounds, also in the presence of a cosmological constant. Plane-waves in flat space were used in \cite{Schroedinger42} as incoming states to test non-linear effects in light-by-light scattering in Born-Infeld's theory. The solutions found in the present contribution could be similarly employed in a more general context, as well as in string theory applications on the line of \cite{HorSte90,Coley02}.

 In addition, some further conclusions can be drawn from our results. First, we observe that a time-dependent scalar field $\varphi(u)$ (e.g., a dilaton) in the background \eqref{metric} automatically solves any equation $\Box\varphi+\ldots=0$, where the dots can describe any scalar invariant constructed out of the Maxwell field \eqref{F_N_coords} and the Riemann tensor (including their derivatives) as well as derivatives of $\varphi$ (as follows using results of \cite{Pravdaetal02,OrtPra16}). This provides one also with a class of universal scalar fields accompanying the universal electromagnetic solutions. Moreover, the universality property can further be extended to the gravitational field, as a subset of metrics \eqref{metric} (i.e., those which are Einstein and with $g^{(1)}_{,\zeta}=0$) represents a class of universal spacetimes -- for these, any curvature-constructed corrections to the Einstein equations vanish identically \cite{HerPraPra17}. In a broader sense, one is thus able to identify universal solutions to somewhat general theories containing gravity, 2-forms and scalar fields. In fact, although for definiteness we focused here on 2-form fields in four spacetime dimensions, our results can be straightforwardly generalized to null $p$-forms in arbitrary dimensions (and likely to non-abelian theories, cf. \cite{Deser75,Coleman77,Guven87}), which are of more direct interest for supergravity and string theory applications. Additional details and other extensions (such as taking into account the effect of backreaction) will deserve a separate investigation -- previous related results in special cases include \cite{Schwinger51,Deser75,Guven87,HorSte90,Coley02}.

Finally, it should be observed that the solutions obtained in this paper do not exhaust the universal class. For example, it is easy to see that, within NLE, Schr\"odinger's observation extends to all Maxwell fields for which $I_1$ and $I_2$ are constant (not necessarily zero), thus describing, for example, uniform electric and magnetic fields (at least in a flat background).  More generally, any covariantly constant $\bF$ is universal, even if not null. However, the latter solutions clearly describe electromagnetic fields physically different from the null ones. Further results for non-null fields will be presented elsewhere.

\section*{Acknowledgments}
%\acknowledgments

We thank Sigbj\o rn Hervik for discussions. 
This work has been supported by research plan {RVO: 67985840} and research grant GA\v CR 13-10042S. 
The stay of M.O. at Instituto de Ciencias F\'{\i}sicas y Matem\'aticas, Universidad Austral de Chile has been supported by CONICYT PAI ATRACCI{\'O}N DE CAPITAL HUMANO AVANZADO DEL EXTRANJERO Folio 80150028.

%
%\bibliography{bibl}
%
%
%\bibliographystyle{elsart-num_mio}

\end{document}